\documentclass{article}
\usepackage{spconf,amsmath,graphicx}
\usepackage{xcolor}
\usepackage{mathtools}
\usepackage{etoolbox}
\usepackage{svg}
\usepackage{csquotes}
\newcommand{\etal}{\textit{et al.}}

\title{Attention Driven Fusion for Multi-Modal Emotion Recognition}
\name{Darshana Priyasad, Tharindu Fernando, Simon Denman, Sridha Sridharan, Clinton Fookes\vspace{-2mm}}

\address{Speech and Audio Research Lab - SAIVT\\Queensland University of Technology, Brisbane, Australia}

\begin{document}
%\ninept
%
\maketitle
%
% ========================= abstract ==========================================

\begin{abstract}

Deep learning has emerged as a powerful alternative to hand-crafted methods for emotion recognition on combined acoustic and text modalities. Baseline systems model emotion information in text and acoustic modes independently using Deep Convolutional Neural Networks (DCNN) and Recurrent Neural Networks (RNN), followed by applying attention, fusion, and classification. In this paper, we present a deep learning-based approach to exploit and fuse text and acoustic data for emotion classification. We utilize a SincNet layer, based on parameterized sinc functions with band-pass filters, to extract acoustic features from raw audio followed by a DCNN. This approach learns filter banks tuned for emotion recognition and provides more effective features compared to directly applying convolutions over the raw speech signal. For text processing, we use two branches (a DCNN and a Bi-direction RNN followed by a DCNN) in parallel where cross attention is introduced to infer the N-gram level correlations on hidden representations received from the Bi-RNN. Following existing state-of-the-art, we evaluate the performance of the proposed system on the IEMOCAP dataset. Experimental results indicate that the proposed system outperforms existing methods, achieving \textcolor{black}{$\textbf{3.5\%}$}\footnote[1]{Note: this has been updated from the ICASSP published version due to a small error. Results have been updated to correct the error, but overall findings are unchanged.} improvement in weighted accuracy. 

\end{abstract}


% ========================= methodology ==========================================

\section{Introduction}

With the advance of technology, Human Computer Interaction (HCI) has become a major research area. Within this field, automatic emotion recognition is being pursued as a means to improve the level of user experience by tailoring responses to the emotional context, especially in human-machine interactions \cite{mohanta2018detection}. However, this remains challenging due to the ambiguity of expressed emotions. An utterance may contain subject dependent auditory clues regarding expressed emotions which are not captured through speech transcripts alone. With deep learning, architectures can extract higher level features and more robust features for accurate speech emotion recognition \cite{tzirakis2018end, tarantino2019self}. In this paper, we propose a model that combines acoustic and textual information for speech emotion recognition.
%chatziagapi2019data, commented from above cite

Recently, multi-modal information has been used in emotion recognition in preference to uni-modal methods \cite{el2011survey}, since humans express emotion via multiple modes \cite{yoon2019speech, yoon2018multimodal, gu2018deep, li2019attentive}. Most state-of-the-art methods for utterance level emotion recognition have used low-level (energy) and high-level acoustic features (such as Mel Frequency Cepstral Coefficients (MFCC) \cite{yoon2019speech, nguyen2018deep}). However, when the emotion expressed through speech becomes ambiguous, the lexical content may provide complementary information that can address the ambiguity. %As an example, a part of speech may contain words that represent specific emotions such as \enquote{wow} for a surprise. 

Tripathi \etal \cite{tripathi2018multi} has used a Long-Short Term Memory (LSTM) along with a DCNN to perform joint acoustic and textual emotion recognition. They have used features such as MFCC, Zero Crossing Rate and spectral entropy for acoustic data, while using Glove \cite{pennington2014glove} embeddings to extract a feature vector from speech transcripts. However, the performance gain is minimal due to the lack of robustness in the acoustic features, and the sparse text feature vectors. Yenigalla \etal \cite{yenigalla2018speech} proposed a spectral and phenoms-sequence based DCNN model which is capable of retaining the emotional content of the speech that is lost when converted to text. Yoon \etal \cite{yoon2019speech} presented a framework using Bidirectional LSTMs to obtain hidden representations of acoustic and textual information. The resultant features are fused with multi-hop attention, where one modality is used to direct attention for the other mode. Higher performance has been achieved due to the attention and fusion which select relevant segments from the textual model, and complementary features from each mode. Yoon \etal \cite{yoon2018multimodal} have also presented an encoder based method where the fusion of two recurrent encoders is used to combine features from audio and text. However, both methods use manually calculated audio features which limit their accuracy and the robustness of the acoustic model \cite{lee2018convolutional, chen20183}.

Gu \etal \cite{gu2018deep} presented a multimodal framework where a hybrid deep multimodal structure that considers spatial and temporal information is employed. Features obtained from each model were fused using a DNN to classify the emotion. Li \etal \cite{li2019attentive} proposed a personalized attribute aware attention mechanism where an attention profile is learned based on acoustic and lexical behavior data. Mirsamadi \etal \cite{mirsamadi2017automatic} used deep learning along with local attention to automatically extract relevant features where segment level acoustic features are aggregated for utterance level emotion representation. However, the accuracy could be further improved by fusing the audio features with another modality with complementary information. 

In this paper, we present a multi-modal emotion recognition model with combines acoustic and textual information by using DCNNs, and both cross attention and self-attention. Experiments are performed on the IEMOCAP \cite{busso2008iemocap} dataset to enable fair comparison with state-of-the-art methods, and a performance gain of $3.5\%$ in weighted accuracy is achieved.

% ========================= methodology ==========================================

\section{Methodology}

\begin{figure*}[th]
\centering
\includegraphics[width=17.5cm]{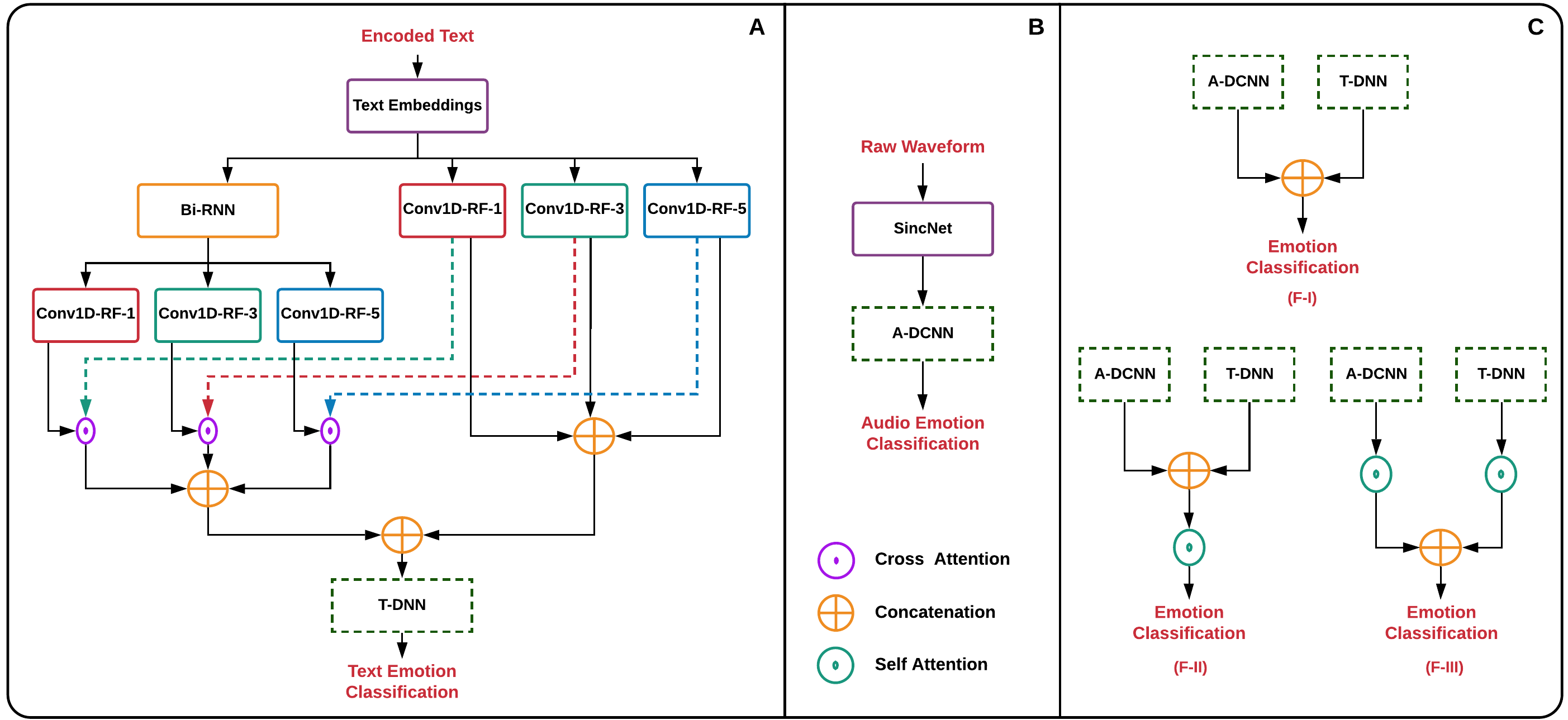}
\caption{Proposed architecture - The system contains three main parts: the text network (A); the audio network (B); and the fusion network (C). The raw audio is passed to the acoustic network after amplitude normalization. It contains SincNet filter layers containing parameterized sinc functions with band-pass filters to extract acoustic features, followed by the A-DCNN component containing convolutional and dense layers. The corresponding text is converted to a word vector using Glove embeddings, and then passed through the text network. It contains two parallel branches with bi-LSTM and convolutional layers (\textcolor{black}{8} filters) with different filter sizes to capture n-grams (n-words) in one iteration where n = \{1,3,5\}. As shown in the figure, convolutional layers in the right branch are used as cross-attention for the left branch. The two branches are fused, followed by the T-DNN component for textual emotion recognition. A-DCNN and T-DNN are then fused using self-attention to get the final emotion classification result.}
  \vspace{-6mm}
  \label{fig:nn}
\end{figure*}

\subsection{Acoustic Feature Extraction}

In our proposed model, we utilize a SincNet filter layer \cite{ravanelli2018speaker} to learn custom filter banks tuned for emotion recognition from speech audio. This layer is shown to have fast convergence and higher interpretability with a smaller number of parameters compared to conventional convolution layers. Formally, this layer can be defined as, 
 \vspace{-1mm}
 \begin{equation}
    y[n] = x[n] * g[n,\theta],
  \label{conv}
\end{equation}
where $x[n],y[n],g,\theta$ refers to the input signal, filtered output, filter-bank function, and the learnable parameters respectively. In SincNet filters, convolution operations are applied over a raw waveform with predefined functions. Each defined filter-bank is composed of rectangular band-pass filters which can be represented by two low-pass filters with learnable cutoff frequencies. The time-domain representation of the function $g$ can be derived as follows \cite{ravanelli2018speaker},
\begin{equation}
g[n,\ f_{1},\ f_{2}] = 2f_{2}\ sinc(\ 2\pi f_{2}n\ ) - 2f_{1}\ sinc(\ 2\pi f_{1}n\ ),
  \label{function}
\end{equation}
where $f_{1},f_{2}$ refers to low and high cutoff frequencies and $sinc = \sin{x}/x$.

The resultant convolution layer outputs are passed through a DCNN which contains several \enquote{Convolution1D}, \enquote{Batch Normalization} and \enquote{fully connected layers}. During the initial training phase, a random $250ms$ chunk from the audio signal is selected as the input. During validation and testing, we obtain the final \enquote{softmax} response for each chunk and add them together to get the final classification scores, similar to \cite{ravanelli2018speaker}. However, we retrieve a 2048-D feature vector from the final dense layer before the classification layer for each chunk of an utterance, and average these before fusing them with textual features from the corresponding transcript in a later step (see Section \ref{subsec:fusion}).

\subsection{Textual Feature Extraction}

In our proposed model, after the sequence vector is passed through a common embedding layer, we utilize two parallel branches for textual feature extraction as illustrated in Figure \ref{fig:nn}. Bi-RNNs followed by DCNNs have been extensively used in textual emotion analysis \cite{etienne2018cnn+, ramet2018context}. As an alternative, a CNN based architecture which is capable of considering \enquote{n} words at a time (n-grams) can be used \cite{kim2014convolutional}. Therefore we use two parallel branches, employing, one using Bi-RNNs with DCNNs and the other DCNNs alone to increase the effectiveness of the learned features (see Figure \ref{fig:nn} (B)). The resultant feature vector from the Bi-RNN is passed through three convolutional layers with a filter sizes of $1$ ,$3$ and $5$; and convolutional layers with the same size filter are used in the parallel branch. We introduce cross-attention where we use convolution layers with the same filter size from the right branch as the attention for the left branch, as illustrated, and jointly train with the other components of the network. The cross-attention is calculated using
\vspace{-1mm}
\begin{equation}
    \alpha_i = \frac{exp((\textbf{b}_{i,j})^{\top}\ \textbf{a}_{i,j}^\top)}{\sum\limits_{i}^{} \ exp((\textbf{b}_{i,j})^{\top}\  \textbf{a}_{i,j}^\top)},
    \label{text1}
\end{equation}
\begin{equation}
    H = \sum\limits_{i}^{} \ \alpha_i\ \textbf{a}_{i,j}^\top,
  \label{text2}
\end{equation}
where $\alpha_i, b_{i,j}, a_{i,j}, H$ are the attention score, context vector from the right branch with a filter size of $j$, output of the convolution layer with filter size $j$ in left branch, and the output.

The convolution layers from both branches are concatenated together and passed through a DNN consisting of fully connected layers for textual emotion classification. We retrieve a  4800-D feature vector from the final dense layer before the classification layer for multi-modal feature fusion.

\subsection{Acoustic \& Textual Feature Fusion}
\label{subsec:fusion}

Mid-level fusion is used to fuse textual and acoustic features obtained from individual networks. A 2048-D feature vector from the acoustic network and a 4800-D feature vector from the textual network are concatenated as illustrated in Figure \ref{fig:nn} (C). A neural network with attention is used to identify informative segments in the feature vectors. We have explored using fusion without attention (F-I), attention after fusion (F-II) where self-attention is applied on concatenated features, and attention before fusion (F-III) where attention is applied on individual feature vectors. 
For F-III, we calculate attention weights and combine the vectors using \cite{priyasadlearning},

\vspace{-1mm}
\begin{equation}
\textcolor{black}{h_t = tanh\:(\:\textbf{W}_h\:\textbf{c}_t\:+\:b_h\:),}
\label{normal1}
\end{equation}

\begin{equation}
\textcolor{black}{\beta_t = \rm{sigmoid} (h_t),}
\label{normal2}
\end{equation}

\begin{equation}
\textcolor{black}{q = \beta_t\:\otimes \textbf{c}_t}
\label{normal3}
\end{equation}

\textcolor{black}{$c_{\text{t}}$, $h_{\text{t}}$, $\beta_{\text{t}}$, and $q$ refer to the merged feature vector, neural network (which is randomly initialized and jointly trained with other components of the network) output, attention score and the output respectively}. Finally the the utterance emotion is classified using a \enquote{softmax} activation over the final dense layer of the fusion network.

% ========================= results ==========================================

\section{Experiments}

\subsection{Dataset and Experimental Setup}

Experiments are conducted on the Interactive Emotional Dyadic Motion Capture (IEMOCAP) dataset which includes five sessions of utterances for 10 unique speakers. We follow the evaluation protocol of \cite{yoon2019speech, yoon2018multimodal}, and select utterances annotated with four basic emotions \enquote{anger}, \enquote{happiness}, \enquote{neutral} and \enquote{sadness}. Samples with \enquote{excitement} are merged with \enquote{happiness} as per \cite{yoon2019speech, yoon2018multimodal}. The resultant dataset contains $5531$ utterances \{\enquote{anger}:$1103$, \enquote{happiness}:$1636$, \enquote{neutral}:$1708$, \enquote{sadness}:$1084$\}. 

Initial training is carried out on both acoustic and textual networks separately before the fusion. The sampling rate of each utterance waveform is set to $16,000$Hz while a random segment of $250ms$ is used in training the acoustic network. During the evaluation, the cumulative sum of all the predictions with a window size and shift of $250ms$ and $10ms$ are considered. In the textual network, all the transcripts of the utterances are set to a maximum length of $100$ and padded with 0s. Glove-300d embeddings are used to convert the word sequence to a vector of $(100,300)$. \textcolor{black}{We utilize a 10-fold cross-validation with an 8:1:1 split for training, validation, and test sets respectively for text model. We select an average performing split and use this split to train the acoustic and fusion networks (we use a single split due to the high computation time of the acoustic model), such that all networks (text, audio and fusion) use the same data splits.} The learning rate and the batch size in each network are fixed at $0.001$ and $64$ respectively, and the Adam optimiser is used. 

\subsection{Performance Evaluation and Analysis}

\begin{figure*}
\centering
\includegraphics[width=17.5cm]{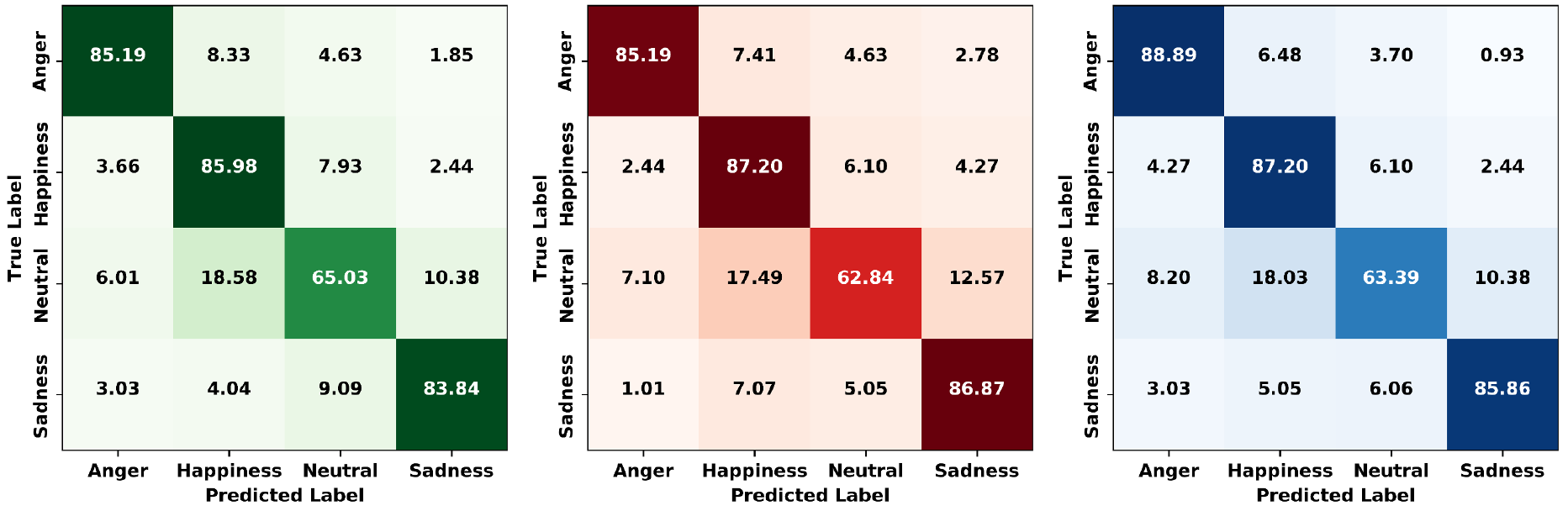}
\caption{Confusion matrices of the proposed architecture for separate fusion methods calculated using a average performing 8:1:1 split. Left, middle and right figures represent Fusion I, Fusion II and Fusion III respectively.}
\label{fig:confusion}
\vspace{-2mm}
\end{figure*}

Following \cite{yoon2019speech, yoon2018multimodal}, the performance of our system is measured in terms of weighted accuracy (WA) and unweighted accuracy (UA). Table \ref{acc} and Figure \ref{fig:confusion} present performance of our approach for emotion recognition compared with the state of the art methods.

MDRE \cite{yoon2018multimodal} has used two RNNs to encode both acoustic and textual data followed by a DNN for classification, while Evec-MCNN-LSTM \cite{cho2018deep} has used an RNN and a DCNN to encode both modalities followed by fusion and an SVM for classification. MHA-2 \cite{yoon2019speech} has used two Bidirectional Recurrent Encoders (BRE) for both modalities followed by a multi-hop attention mechanism. MDRE has outperformed MCNN-LSTM by $10.6\%$ and MHA-2 has outperformed MDRE by $6.5\%$, demonstrating how attention can increase performance.

Our proposed model has achieved a substantial improvement in overall accuracy, with a $3.5\%$ increase compared to MHA-2. We have utilized self-attention before (F-III) and after fusion (F-II) as illustrated in Figure \ref{fig:nn}. Cross-modal attention has not been utilized after fusion since the dimensionality of the feature vectors from the two modalities are different. A slight increase in the classification accuracy has been obtained by applying self-attention compared to conventional feature fusion (F-I). Furthermore, the highest accuracy has been obtained by \textbf{F-III}, outperforming F-II by $0.5\%$. Given that F-I slightly outperforms MHA-2, we have compared the classification accuracy of the individual modes of MHA-2 with our individual modes in Table \ref{acci}.

\begin{table}[t]
\centering
\begin{tabular}{|c|c|c|c|c|} 
\hline 
Model & Modality & WA & UA \\ [0.5ex]
\hline \hline
Evec-MCNN-LSTM \cite{cho2018deep}& $A+T$ & $64.9\%$ & $65.9\%$\\
MDRE \cite{yoon2018multimodal}& $A+T$ & $71.8\%$ & $-$\\
MHA-2 \cite{yoon2019speech}& $A+T$ & $76.5\%$ & $77.6\%$\\
\hline
\textbf{Ours - F-I} & $A+T$ & $\textcolor{black}{77.85\%}$ & $\textcolor{black}{79.27\%}$\\
\textbf{Ours - F-II} & $A+T$ & $\textcolor{black}{78.98\%}$ & $\textcolor{black}{80.01\%}$\\
\textbf{Ours - F-III} & $A+T$ & $\textcolor{black}{\textbf{79.22\%}}$ &$\textcolor{black}{\textbf{80.51\%}}$\\
\hline
\end{tabular}
\caption{\textcolor{black}{Recognition accuracy for IEMOCAP using an average performing 8:1:1 split, compared with state of the art methods.}}

\label{acc}
\end{table}

\begin{table}[t]
\centering
\begin{tabular}{|c|c|c|} 
\hline 
Model & Modality & WA \\ [0.5ex]
\hline \hline
MHA-2 \cite{yoon2019speech}& $A$  & $65.2\%$\\
MHA-2 \cite{yoon2019speech}& $T$  & $70.3\%$\\
\hline
\textbf{Ours} & $A$  & \textcolor{black}{$\textbf{69.8\%}$}\\
\textbf{Ours} & $T$  & \textcolor{black}{$\textbf{66.7\%}$}\\
\hline
\end{tabular}
\caption{\textcolor{black}{Recognition accuracy of individual modes of the IEMOCAP dataset with an average performing 8:1:1 split, comparing the proposed approach with \cite{yoon2019speech}.}}
\label{acci}
\end{table}

Our acoustic and textual models outperformed the corresponding individual modes of MHA-2 \cite{yoon2019speech}, where a substantial improvement of $\textbf{7.2\%}$ is achieved with the acoustic model. The SincNet layer in the acoustic model is capable of learning and deriving customized filter banks tuned for emotion recognition. It has been successfully applied for speaker recognition \cite{ravanelli2018speaker} as an alternative to i-vectors. The confusion matrices for F-I, F-II, and F-III are illustrated in Figure \ref{fig:confusion}. A $\textbf{10.6\%}$ relative improvement in classification accuracy can be observed when comparing the individual modalities with the fusion network in our model. Given that accuracy is approximately similar for both modalities, each modality has complemented the other to increase recognition accuracy. 

\vspace{-5mm}

% ========================= conclusion ==========================================
% \vspace{-3mm}
\section{Conclusion}

In this paper, we present an attention-based multi-modal emotion recognition model combining acoustic and textual data. The raw audio waveform is utilized in our method, rather than extracting hand-crafted features as done by baseline methods. Combining a DCNN with a SincNet layer, which learns suitable filter parameters over the waveform for emotion recognition, outperforms the hand-crafted feature-based audio emotion detection of the baselines. Cross attention is applied to text-based feature extraction to guide the features derived by RNNs using N-gram level features extracted by a parallel branch. We have used self-attention on both feature vectors obtained from two networks before the fusion, to attend to the informative segments from each feature vector. We have achieved a weighted accuracy of \textcolor{black}{$\textbf{79.22\%}$} on the IEMOCAP database, which outperforms the existing state-of-the-art model by \textcolor{black}{$\textbf{3.5\%}$}.

%
% ========================= acknowledgemet ==========================================

\section{Acknowledgements}

This research was supported by an Australia Research Council (ARC) Discovery grant DP140100793.

\bibliographystyle{IEEEtran}

\end{document}